\documentclass[12pt]{article}
\usepackage{graphicx}
\usepackage{amssymb}
\usepackage{bm}
	
\renewcommand{\abstract}[1]{{ \footnotesize \noindent {\bf Abstract} #1 \\}}
\renewcommand{\author}[1]{\subsubsection*{\it#1}}
\newcommand{\address}[1]{\subsubsection*{\it#1}}
\def\lsim{\lower.5ex\hbox{$\; \buildrel < \over \sim \;$}}
\def\gsim{\lower.5ex\hbox{$\; \buildrel > \over \sim \;$}}

\newcommand{\chapter}[1]{{\Large \bf \noindent #1}}

\begin{document}
\chapter{Nucleosynthesis in stellar flares}

\author{V. Tatischeff $^{1,2}$, J.-P. Thibaud $^1$, I. Ribas $^2$}

\address{$^1$Centre de Spectrom\'etrie Nucl\'eaire et de Spectrom\'etrie de 
Masse, CNRS/IN2P3 and Univ Paris-Sud, F-91405 Orsay, France\\
$^2$Institut de Ci\`encies de l'Espai (CSIC-IEEC), Campus UAB, 
Fac. Ci\`encies, 08193 Bellaterra, Barcelona, Spain}

   \abstract{Nuclear interactions of ions accelerated at the surface of flaring 
   stars can produce fresh isotopes in stellar atmospheres. Although this nucleosynthesis is
   not significant for the chemical evolution of the Galaxy, it can be important for a number
   of measurements of ``anomalously'' high $^6$Li and $^7$Li abundances. We discuss the 
   possible role of stellar flares to explain the recent report of high $^6$Li 
   abundances in metal-poor halo stars and the well-established correlation between Li 
   abundance and stellar activity in young open clusters. We then study the possibility
   of observing directly Li production during flares of nearby and active dwarfs of spectral 
   type M.} 
%

\section{Introduction}
\label{secvt1}

Nuclear interactions of ions accelerated in solar flares produce a characteristic 
gamma-ray emission in the solar atmosphere. Repeated spectroscopic observations of 
this emission have furnished valuable information on the physical conditions prevailing 
within solar active regions, as well as on the composition, energy spectrum, and angular
distribution of the flare-accelerated ions \cite{mur07}. The solar-flare 
gamma-ray line emission testifies that fresh nuclei are synthesized in abundance in
energetic solar events. Thus, the gamma-ray lines at 478 and 429~keV emitted in the 
reactions $^4$He($\alpha$,$p$)$^7$Li and $^4$He($\alpha$,$n$)$^7$Be, respectively,
allow to trace the production of $^7$Li ($^7$Be decays to $^7$Li 
with a half-life of 53~days). There is no observable gamma-ray emission 
associated with the synthesis of $^6$Li. But evidence for significant production of 
this isotope in large solar flares is provided by optical observations of sunspots 
\cite{rit97} and measurements of the solar wind Li isotopic ratio in lunar soil 
\cite{cha99}. 

Solar-type activity is believed to be a phenomenon inherent to the vast majority if 
not all main-sequence stars. The Sun is not an active star in comparison with
numerous stellar objects in the solar neighbourhood that show much higher
luminosities in emissions associated with coronal and chromospheric activities. Although
gamma-ray line emission from other flaring stars cannot be observed with the current 
sensitivity of the gamma-ray space instruments, it is more than likely that the Sun is not 
the only star producing surface nucleosynthesis in flares. 

It is instructive to compare the energetics of stellar flares with that of the galactic
cosmic rays, as the latter are a significant source of nucleosynthesis of the light 
elements Li, Be and B \cite{ree70}. There are typically 375 gamma-ray flares per solar 
cycle (duration 11~yr) each releasing on average about 10$^{31}$~erg of kinetic energy in 
accelerated ions of energy $\geq$1~MeV~per~nucleon \cite{ram00}. The power in 
these particles averaged over a solar cycle is $\sim$10$^{25}$~erg~s$^{-1}$. Assuming 
that there are about 10$^{11}$ flaring stars with the solar activity in the Milky way, one 
gets a total power of $\sim$10$^{36}$~erg~s$^{-1}$ available for stellar-flare 
nucleosynthesis, which is about 10$^5$ times less than the power contained in the galactic 
cosmic rays. This shows that nucleosynthesis in stellar flares is not important for the 
galactic chemical evolution. However, it can produce fresh nuclei {\it in situ}, i.e. in
stellar atmospheres where abundances are usually measured. The purpose of this paper is to 
study the possible role of stellar flares in a few selected abundance measurements. 

In the next section, we first calculate the yields for synthesis of the light elements 
in stellar flares and show that the Li isotopes should generally be more efficiently 
produced than the other species. In Sect.~\ref{secvt3} we discuss the recent report of 
relatively high $^6$Li abundances in metal-poor halo stars. In Sect.~\ref{secvt4} we 
propose a new model based on stellar flare nucleosynthesis to explain the observed 
correlation between Li abundance and stellar activity in young open clusters. In 
Sect.~\ref{secvt5} we study the possibility of observing directly the creation of Li 
during large flares of active M-type dwarfs. Conclusions are given in 
Sect.~\ref{secvt6}. 

\section{Yields of light element production in stellar flares}
\label{secvt2}

The light elements Li, Be and B are a priori the best candidates to trace nonthermal
nucleosynthesis in stellar flares, because heavier species are presumably produced in too
large amounts by thermal nucleosynthesis in stellar interiors and explosions. To calculate 
the efficiency for light element production in stellar flares, we assume the same thick 
target interaction model that is usually employed to describe nuclear processes in solar 
flares (see \cite{tat06}). Fast particles with given energy spectra and composition are 
supposed to be accelerated in the stellar corona and to produce nuclear reactions as they 
slow down in the lower  atmosphere. We use for the source energy distribution of the fast 
ions a power law of spectral index $s$. Analyses of the gamma-ray line emission produced 
in solar flares have shown that the spectral index is usually in the range $3<s<5$. We 
assume that $s \simeq 4$ is also close to the mean of spectral index distribution in 
stellar flares. 

The ambient and accelerated-ion abundances are also based on those employed in solar flare 
studies \cite{tat06}, but we rescale the abundances of both ambient and fast C and heavier 
elements to the metallicity of the star we are studying. We adopt 
an accelerated $\alpha$-to-proton abundance ratio of 0.1. Enormous enhancements of
accelerated $^3$He are measured in impulsive solar flares: the $^3$He/$\alpha$ ratios
found in these events are frequently three to four orders of magnitude larger than the
corresponding value in the solar corona and solar wind, where 
$^3$He/$^4$He$\sim$$5\times10^{-4}$ \cite{rea99}. Such an enrichment of accelerated $^3$He 
is caused by resonant wave-particle processes that are characteristic of the stochastic 
acceleration mechanism at work in impulsive solar flares. It is reasonable to assume that 
this acceleration process enhances the accelerated $^3$He in stellar flares as well, but
the abundance of fast $^3$He is nonetheless uncertain. 

The cross sections for the nuclear reactions producing the light elements are mostly from 
Ref.~\cite{ram97}, but we took into account the more recent measurements of 
Ref.~\cite{mer01} for production of $^6$Li and $^7$Li in the $\alpha + \alpha$ reactions 
and the evaluation of Ref.~\cite{ram00} for the cross section of the reaction 
$^4$He($^3$He,$p$)$^6$Li. 
 
   \begin{figure}[t]
   \centering
   \includegraphics[width=9.cm]{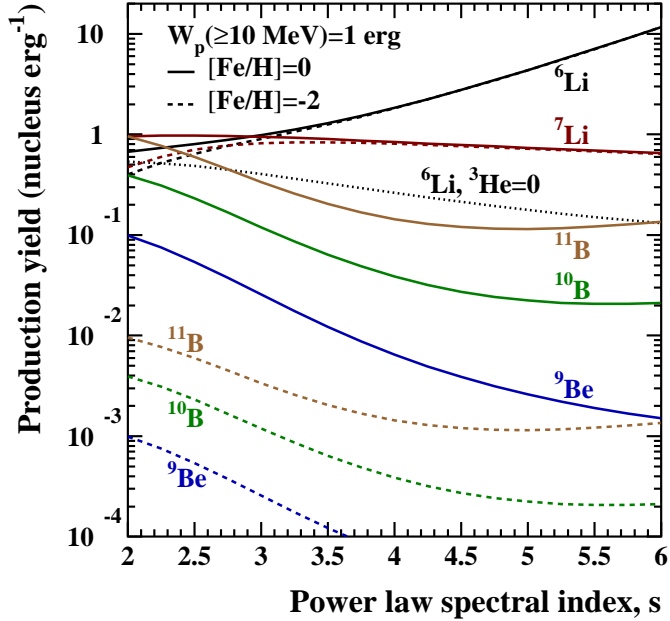}
      \caption{Yields of light element production in stellar flares as a function of the 
      power-law spectral index of the accelerated particle energy distribution. The 
      calculations are normalized to a total kinetic energy of 1 erg in protons of energy 
      $E\ge10~{\rm MeV}$. {\it Solid curves}: [Fe/H]=0 and accelerated $^3$He/$\alpha$=0.5 
      {\it Dashed curves}: [Fe/H]=-2 and $^3$He/$\alpha$=0.5. {\it Dotted curve}: $^6$Li 
      production for [Fe/H]=0 and $^3$He=0. 
              }
         \label{figvt1}
   \end{figure}

Figure~\ref{figvt1} shows calculated production yields as a function of $s$. The 
calculations are normalized to a total kinetic energy of 1 erg contained in 
flare-accelerated protons of energy greater than 10~MeV. We see that for [Fe/H]$\leq$0 
([Fe/H]=$\log[{\rm (Fe/H)/(Fe/H)}_\odot]$ and (Fe/H)/(Fe/H)$_\odot$ is the Fe abundance 
relative to its solar value) the Li isotopes are the most 
efficiently synthesized and that for $s>3$ and $^3$He/$\alpha=0.5$, the largest 
production is that of $^6$Li. The importance of 
the reaction $^4$He($^3$He,$p$)$^6$Li for the synthesis of this isotope can be seen by 
comparing the upper curve with the dotted one, for which we set the accelerated $^3$He 
abundance to zero. We find that for $s=4$ and $^3$He/$\alpha=0.5$, the $^3$He+$^4$He 
reaction accounts for 85\% of the total $^6$Li production. 

The metallicity dependence of the production yields can be seen by comparing 
the curves obtained for [Fe/H]=0 with those for [Fe/H]=-2. The productions of
Be and B are proportional to the abundance of metals, because these species 
result from spallation of fast (resp. ambient) C, N and O interacting with
ambient (resp. fast) H and He. On the other hand, the productions of $^6$Li and 
$^7$Li are almost independent of metallicity for $s>3$, because the Li
isotopes are then produced almost exclusively in accelerated $^3$He and 
$\alpha$-particle interactions with ambient He. 

\section{$^6$Li production in flares of metal-poor halo stars}
\label{secvt3}

Asplund et al. \cite{asp06} have recently reported the detection of $^6$Li at 
$\geq$2$\sigma$ confidence level in nine halo stars of low metallicity, [Fe/H]$<$-1, 
situated in the turnoff region of the Hertzsprung-Russel diagram. The $^6$Li 
abundances measured in these objects are far above the value predicted by 
Big Bang nucleosynthesis and cannot be explained by galactic cosmic-ray interactions 
in the interstellar medium either. Proposals to explain these observations 
include:  $^6$Li production in the early universe induced by the decay of 
supersymmetric dark matter particles during Big Bang nucleosynthesis, (2) production 
by the interaction of cosmological cosmic rays that could be accelerated in shocks 
induced by large-scale structure formation or by an early population of massive stars
(Pop~III stars), and (3) variations of the fundamental physical constants. However, 
the $^6$Li abundances reported in Ref.~\cite{asp06} and previous works are now questionned, 
as Cayrel et al. \cite{cay07} have just shown that convective Doppler shifts in stellar 
atmospheres can generate an asymmetry of the Li I line at 6707.8~\AA~similar to the slight 
blending produced by the contribution of $^6$Li.

Tatischeff \& Thibaud \cite{tat07} have shown that a significant amount of $^6$Li can
be produced in the atmospheres of metal-poor halo stars from repeated solar-like flares
during their main-sequence evolution. These authors have developed a model of light 
element production by stellar flares, which is based on our current knowledge of the 
flaring activity of main-sequence dwarf stars. It is now well established that stellar
activity depends primarily on two quantities: the effective temperature, which is the
best indicator of how deep is the surface convection zone of the star, and the 
rotation rate, which controls the generation and amplification of magnetic fields by 
a complex $\alpha-\Omega$ process in the convective envelope. In the model of
Ref.~\cite{tat07}, the decisive factor for a significant flare production of $^6$Li is 
the star rotation period at the zero-age main sequence, $P_{\rm ZAMS}$. The flare 
productions of the other light species $^7$Li, Be and B were found to be always negligible 
as compared with the observed abundances of these elements in metal-poor halo stars. 

   \begin{figure}
   \centering
   \includegraphics[width=7.5cm]{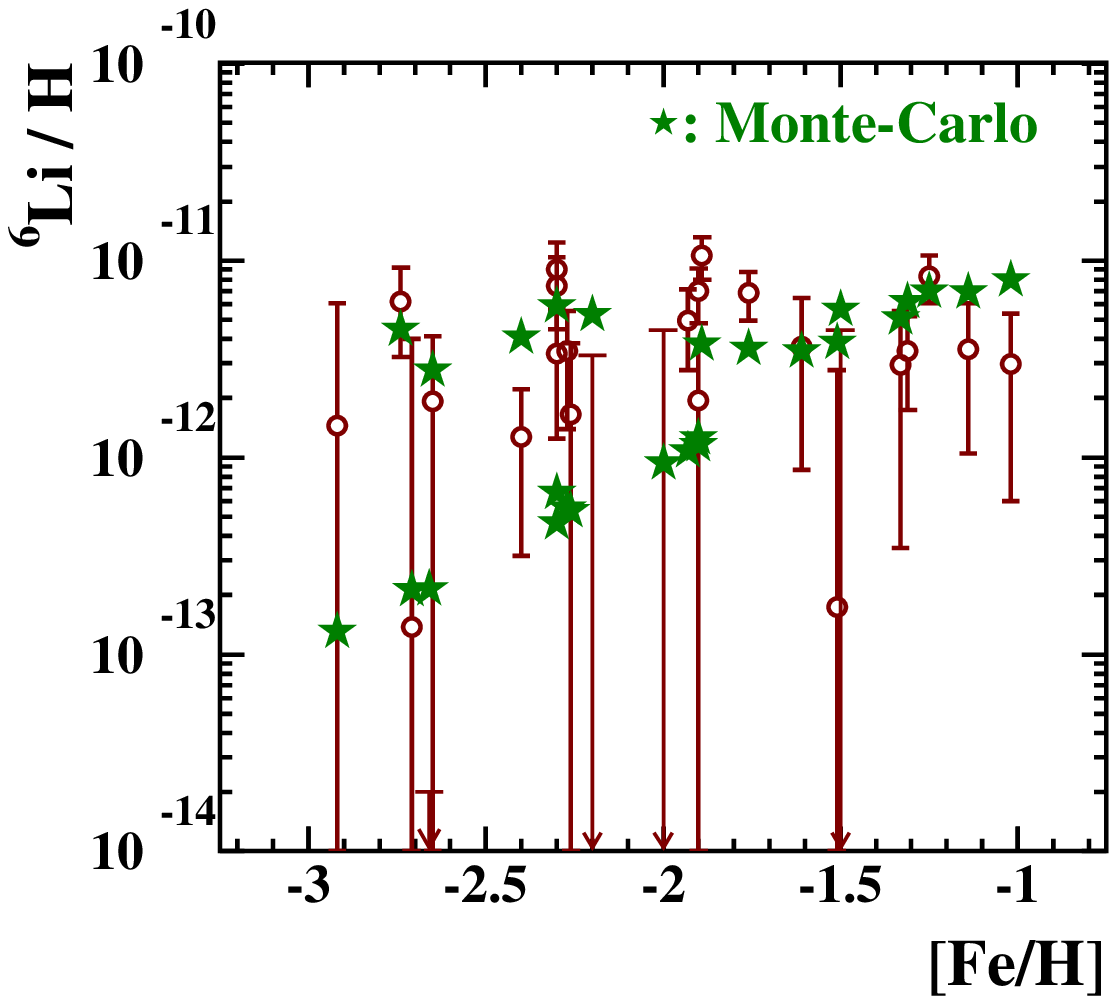}
   \includegraphics[width=7.5cm]{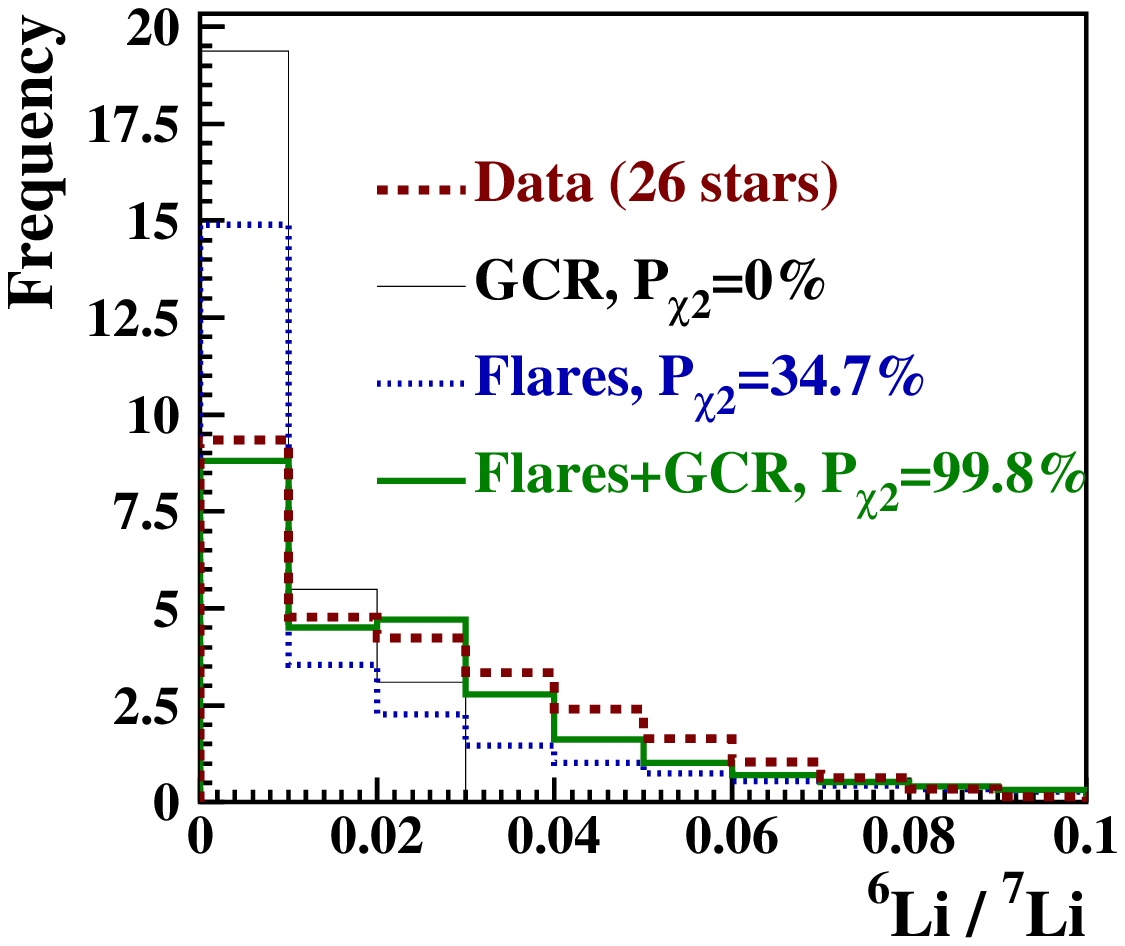}
      \caption{{\it Left panel}: $^6$Li surface abundances in metal-poor halo stars 
as a function of metallicity. The results of a Monte-Carlo draw of 26 stars from the 
flare model of Ref.~\cite{tat07} is compared with the data of Ref.~\cite{asp06} 
and references therein. {\it Right panel}: calculated and observed $^6$Li/$^7$Li 
ratios averaged over metallicity. The values of $P_{\chi^2}$ indicated in the
figure give the probability that the $^6$Li/$^7$Li data sample comes from a
specific distribution calculated in the framework of the model (see 
\cite{tat07}). GCR: galactic cosmic ray.
              }
         \label{figvt2}
   \end{figure}

We show in Fig.~\ref{figvt2} a comparison of the $^6$Li data of Ref.~\cite{asp06} and
references therein with a Monte-Carlo simulation \cite{tat07} in which stars were 
generated at random with constraints on their mass, metallicity and $P_{\rm ZAMS}$. 
The statistical distribution of $P_{\rm ZAMS}$ has been obtained from the study of 
Herbst \& Mundt \cite{her05}, who found that 50--60\% of young main-sequence dwarf 
stars are rapidly rotating ($P_{ZAMS}$$\lsim$2~days), probably because they were 
released from any accretion disk locking mechanism very early on and thus conserved 
angular momentum throughout most of their pre-main sequence evolution, whereas  the 
remaining 40--50\% stars that are more slowly rotating lost substantial amounts of 
angular momentum during their first million years. The effect of this bimodal 
distribution on the calculated $^6$Li production can be seen in the left panel of 
Fig.~\ref{figvt2}. For [Fe/H]$<$-1.7, about half of the simulated stars have 
$^6$Li/H$>$$2\times10^{-12}$, as a result of a significant $^6$Li production by flares 
in these stars that have $P_{ZAMS}$$\lsim$2~days. The lower $^6$Li abundances of the other
stars that are slow rotators mainly come from the simulated galactic cosmic-ray (GCR) 
nucleosynthesis, as the flare production was found to be very small in these objects. 
For [Fe/H]$>$-1.7, the flare and GCR contributions are of similar magnitude. 

The $^6$Li/$^7$Li frequency distribution measured in metal-poor halo stars can result from 
a combination of flare-produced $^6$Li with a protostellar $^6$Li component from GCR 
nucleosynthesis (see Fig.~\ref{figvt2}, right panel). This scenario implies a relatively 
large dispersion of the $^6$Li abundances at low metallicities, which is compatible with 
the current data (see \cite{tat07}). It predicts that the metal-poor stars with the
highest $^6$Li abundances should have higher rotation velocities.

\section{The correlation between Li abundance and stellar activity in 
young open clusters}
\label{secvt4}

\begin{figure}[t]
   \centering
   \includegraphics[width=15.cm]{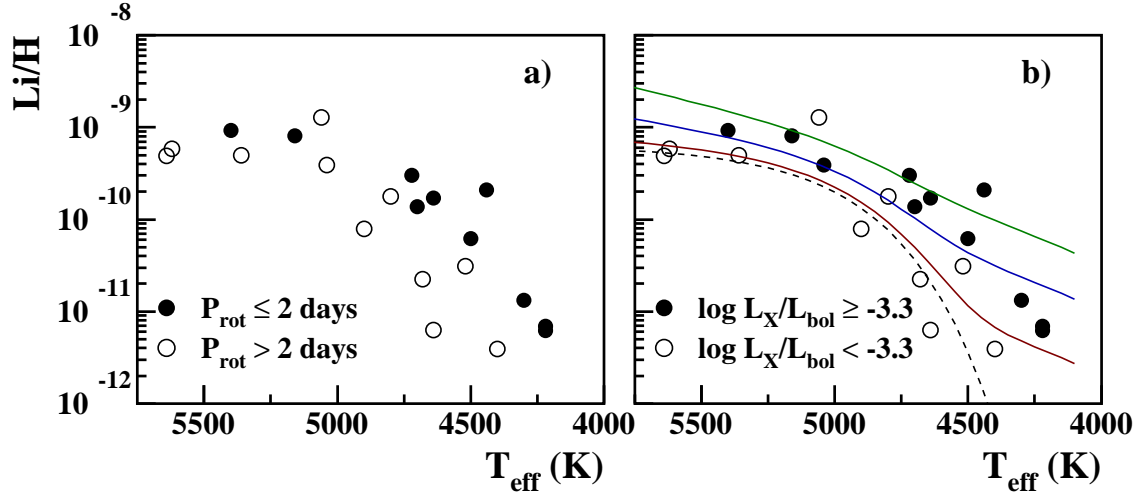}
      \caption{Li abundances as a funcation of effective temperature for stars
of the Pleiades. (a) {\it Filled circles}: $P_{\rm rot} \leq 2$~days; {\it open 
circles}: $P_{\rm rot} > 2$~days. (b): {\it Filled circles}: 
$\log (L_X/L_{\rm bol}) \geq -3.3$; {\it open circles}: $\log (L_X/L_{\rm bol}) 
< -3.3$. {\it Dashed curve}: calculated Li/H without flare production; 
{\it solid curves}: total Li abundances where the flare Li production is
obtained from eq.(\ref{eqvt4}) with $\alpha=1.5$ and $\mu=0.5$, and for three 
values of $\log (L_X/L_{\rm bol})$: -2.8 ({\it top}), -3.3 ({\it middle}) and -4 
({\it bottom curve}). The Li/H and $T_{\rm eff}$ data are from \cite{sod93}, 
and the $P_{\rm rot}$ and $L_X/L_{\rm bol}$ data are from \cite{piz03} and 
references therein. The $L_X$ values were measured in the {\it 
ROSAT}/PSPC 0.1--2.4~keV energy band.
              }
         \label{figvt4}
   \end{figure}

A significant star-to-star scatter of the Li abundance for the same $B-V$ color
is observed in all young open clusters for stars with masses lower than $\sim0.9~M_\odot$ 
(e.g. IC~2602 at age 30~Myr, IC~4665 at 35~Myr, $\alpha$ Persei at 50~Myr, the Pleiades at 
70--100~Myr; e.g. \cite{ran98}). The Li/H ratios measured in these clusters were found to 
depend on stellar rotation and activity: the most rapid rotators, which are also the most 
active stars in chromospheric and coronal emissions, appear to be the most Li rich. This is 
illustrated in Fig.~\ref{figvt4} for a selection of Pleiades stars. We see that for a given 
effective temperature, stars of rotation period $P_{\rm rot} \leq 2$~days and 
X-ray-to-bolometric luminosity ratio $L_X/L_{\rm bol} \geq 5 \times 10^{-4}$ have generally 
higher Li abundances. A noticeable exception is the moderately active ($L_X/L_{\rm bol} =
3.7\times 10^{-4}$) star HII~263, which has ${\rm Li}/{\rm H}=1.3\times 10^{-9}$. We also 
note the strong decrease of Li/H for $T_{\rm eff} < 5000$~K, which is due to the rapid 
destruction of Li in K-type stars. In slightly older open clusters like the Hyades and 
Praesepe, both of age $\approx$700~Myr, the observed Li/H dispersion is significantly less 
pronounced, although it is nonetheless inconsistent with the observational errors 
\cite{tho93}. 

The Li--activity correlation is not well understood. It has been suggested that the 
observed dispersion is not due to an intrinsic scatter in Li abundances, but caused by 
observational effects associated with chromospheric activity, like the presence of 
starspots and faculae (e.g. \cite{rus96,xio05}). However, Barrado y Navascu\'es et al. 
\cite{bar01} have found that only a small part of the Li dispersion observed in Pleiades 
stars can be explained by such systematic effects. Recently, Leone \cite{leo07} has 
proposed that the magnetic intensification of Li lines can result in an overestimation of 
Li abundances in active stars, thus contributing to the observed spread.

Garc\'ia L\'opez et al. \cite{gar94} and Randich et al. \cite{ran98} have suggested that 
the connection between Li abundance and rotation could be caused by the inhibition of Li
depeletion in the most rapid rotators during their pre-main sequence evolution. In this
scenario, fast rotators that have dissipated their circumstellar disk at early stages of
their evolution, would have depleted less Li because they have undergone little angular 
momentum loss and transport, and hence little rotationally-driven mixing, until they have 
reached the zero-age main sequence. However, recent observations of weak lined T Tauri 
stars show that rapid rotation does not inhibit Li depletion among low-mass pre-main 
sequence stars \cite{xin07}. This scenario is also not supported by stellar models of
rotation-induced Li depeletion, which predict on the contrary that fast rotation enhances
the mixing processes that lead to Li destruction \cite{cha07,pia02}. 

In this paper, we assess the possibility that the observed Li--rotation correlation is 
due to a significant in situ production of Li by stellar flares in the most active  
main sequence stars. We assume that the Li atoms produced by nonthermal reactions in the
atmosphere of a given star are mainly evacuated by the stellar wind on a relatively short 
timescale, rather than being mixed into the bulk of the star convection zone. Comparison of
the solar wind $^6$Li abundance with calculations of the production of this isotope in
solar flares has shown that this assumption is reasonable for the contemporary Sun
\cite{ram00}\footnote{In \S~\ref{secvt3}, we have assumed on the contrary that 
flare-produced $^6$Li atoms mainly accumulated in the convection zone of the studied
metal-poor halo stars. This assumption was justified by the fact that the mass-loss rates of 
these low-metallicity objects are expected to be very low. Indeed, stellar winds are driven 
by radiation pressure and it is well known that the main source of radiation opacity is 
provided by metal lines.}. At steady state between Li production by flares and loss in the 
wind, the surface Li abundance of a given star can be expressed as
\begin{equation} 
{{\rm Li} \over {\rm H}} = \bigg({{\rm Li} \over {\rm H}}\bigg)_0 + {1.4m_p
\dot{N_f}({\rm Li}) \over \dot{M}} ~,
\label{eqvt1}
\end{equation}
where (Li/H)$_0$ is the surface abundance of Li of protostellar origin, $m_p$ is 
the proton mass, $\dot{N_f}({\rm Li})$ is the Li flare-production rate and 
$\dot{M}$ is the stellar mass-loss rate. Following \cite{tat07}, the Li production 
rate can be estimated to be
\begin{equation} 
\dot{N_f}({\rm Li})  = Q({\rm Li}) 
L_p^\odot(\ge10~{\rm MeV}) \bigg({L_X \over L_X^\odot}\bigg)^\alpha~,
\label{eqvt2}
\end{equation}
where $Q({\rm Li})$ is the Li production yield plotted in Fig.~\ref{figvt1} 
($Q({\rm Li})=2 \pm 1$~atoms per erg for $s=4$, depending on the accelerated 
$^3$He abundance), $L_p^\odot(\ge10~{\rm MeV})$ $\sim$ 10$^{23}$~erg~s$^{-1}$ is 
the average power contained in solar-flare accelerated protons of kinetic energy 
$E$$\ge10~{\rm MeV}$ \cite{ram00}, $L_X$ is the star X-ray luminosity in the {\it 
ROSAT}/PSPC band, $L_X^\odot \sim 10^{27}$~erg~s$^{-1}$ is the 
average solar luminosity in the same X-ray band, and the index $\alpha$ accounts 
for the fact that accelerated proton luminosity may not scale linearly with X-ray 
luminosity, because most of the energetic protons could be produced by the most 
powerful flares, whereas heating of stellar coronae could essentially be related 
to less powerful but more frequent flares. In \cite{tat07}, based on the work of 
\cite{fei02}, we adopted $\alpha=1.5$. 

The mass-loss rates of cool main-sequence stars are poorly known, because the
tenuous and highly ionized winds of these stars cannot be directly detected. 
However, theory predicts that mass loss is associated with coronal heating, such 
that the mass-loss rate should scale with the X-ray luminosity (e.g. \cite{cra07}). 
The former quantity can then be written as:
\begin{equation} 
\dot{M} = \dot{M}_\odot \bigg({L_X^\odot \over L_{\rm bol}^\odot}\bigg)^{-\mu} 
\bigg({L_X \over L_{\rm bol}}\bigg)^\mu ~,
\label{eqvt3}
\end{equation}
where $\dot{M}_\odot = 2.07 \times 10^{-14}~M_\odot$~yr$^{-1}$ is the solar mass 
loss rate \cite{hol07}, $L_{\rm bol}^\odot=3.85\times 10^{33}$~erg~s$^{-1}$ is the
solar bolometric luminosity and the index $\mu$ is related to the fact that coronal
X-ray emission mainly comes from closed magnetic loops, whereas mass loss proceeds 
along open magnetic flux tubes. Holzwarth \& Jardine \cite{hol07} recently estimated 
that $\mu \approx 0.5$, whereas Wood et al. (\cite{woo05} and references therein) 
previously found $\mu$ ranging from $\approx$1 to 1.3. Inserting eqs.~(\ref{eqvt2}) 
and (\ref{eqvt3}) into eq.~(\ref{eqvt1}) we obtain:
\begin{equation} 
{{\rm Li} \over {\rm H}} = \bigg({{\rm Li} \over {\rm H}}\bigg)_0 + {2.7 \times 
10^{-3} \over (3.8\times 10^6)^\mu} \bigg({L_X \over L_{\rm bol}}\bigg)^{\alpha-\mu}
\bigg({L_{\rm bol} \over L_{\rm bol}^\odot}\bigg)^\alpha~.
\label{eqvt4}
\end{equation}

Calculated Li abundances are shown in Fig.~\ref{figvt4}b for $\alpha=1.5$ and 
$\mu=0.5$. We modeled (Li/H)$_0$ as a function of $T_{\rm eff}$ with a simple empirical 
expression of exponential decay form. We see that the flare contribution to the total Li 
abundance can be significant for active stars with saturated X-ray emission, 
$L_X /L_{\rm bol} \sim 10^{-3}$. In particular, in situ production by 
flares can explain the non-negligible amounts of Li detected in Pleiades stars  
with $T_{\rm eff}$ lower than $\sim$4500~K, for which depletion of protostellar Li is 
expected to be $\gsim 3$~dex. We also see that the Li abundances in very active stars with 
$T_{\rm eff}$ greater than $\sim$5600~K can exceed the presently cosmic, i.e. meteoritic 
value ${\rm Li}/{\rm H}=2\times 10^{-9}$ \cite{lod03}. However, this prediction 
crucially depends on the uncertain parameters $\alpha$ and $\mu$. In particular, the 
model is not valid if $\mu$ is significantly larger than 0.5. We also did not take 
into account in these calculations the possible dependence of Li depletion with stellar 
rotation. According to recent stellar evolution models with internal gravity waves
\cite{cha07}, protostellar Li could be slightly more depleted in rapidly rotating stars. 

We find that creation of Li by flares can explain the dispersion in Li abundances 
observed in young open clusters like the Pleiades and $\alpha$ Persei. Future
measurements of the isotopic ratio $^6$Li/$^7$Li could allow to test this possibility.
$^6$Li of protostellar origin should be strongly depleted and therefore undetectable in 
stars with ${\rm Li}/{\rm H} \lsim 2\times 10^{-10}$, because $^6$Li is more rapidly
consumed than $^7$Li in stellar interiors. On the other hand, significant $^6$Li can
be produced by flare-accelerated $^3$He and $\alpha$-particle interactions with 
ambient $^4$He. The Li isotopic ratio from flare production is predicted to be in 
the range $0.3<^6$Li/$^7$Li$<2.2$ for $s=4$ and $0<^3$He/$\alpha<0.5$ 
(Fig.~\ref{figvt1}). 

It is remarkable that this simple model of Li production by stellar flares is also
consistent with the lower dispersion in Li/H observed in intermediate-age and old 
open clusters. The X-ray luminosities of dwarf stars in open clusters and the solar 
neighbourhood are found to rapidly decline with the stellar age $t$ for 
$t \gsim 0.1$~Gyr. Cranmer \cite{cra07} uses the following relationship to describe 
a large selection of X-ray luminosities in the {\it ROSAT}/PSPC band: 
\begin{equation} 
{L_X \over L_{\rm bol}}= {4.48\times 10^{-4} \over 1+(12.76t_{\rm Gyr})^{1.79}}~,
\label{eqvt5}
\end{equation}
where $t_{\rm Gyr}$ is the stellar age in units of Gyr. This gives for the age of the 
Hyades ($t \approx 0.7$~Gyr) a reduction of $L_X/L_{\rm bol}$ by a factor of $\sim$25 as 
compared with the average ratio for the Pleiades. Equation~(\ref{eqvt4}) shows that a 
similar reduction can be expected for the Li dispersion, in good agreement with the Hyades 
observations \cite{tho93}.

\section{Can Li production be observed live during stellar flares?}
\label{secvt5}

There are a number of observations reported in the literature that suggest that a 
significant amount of Li can be synthesized in large stellar flares. For example, 
a relatively high Li abundance has been reported in the K7 flaring star HD~358623 
\cite{mat95}. This rapidly rotating and very active dwarf has later been identified as a 
member of the $\sim$12~Myr-old $\beta$ Pictoris Moving Group \cite{kai04}. Anomalously 
high Li abundances have been measured in other K-type stars showing strong chromospheric 
activity. The possibility that the excess Li was produced in situ by flares has already 
been discussed by Pallavicini et al. \cite{pal92}. $^6$Li enhancement has been found 
during a long-duration flare of a chromospherically active binary \cite{mon98} and in the 
atmosphere of a single chromospherically active K-type dwarf \cite{chr05}. In both cases, 
a flare production has been found to be consistent with the activity level of the object.  

We study here the possibility of observing directly Li synthesis during flares of active 
M-type dwarfs. Li observations in clusters and associations indicate that these cool stars 
($T_{\rm eff} < 4000$~K) fully deplete their initial Li very rapidly, mainly 
during their pre-main sequence evolution. M-type main-sequence stars are almost free of 
Li of protostellar origin. There are several very active M-type dwarfs 
with high X-ray luminosities in the solar neighbourhood, e.g. AD Leo 
($L_X=10^{28.95}$~erg~s$^{-1}$) and EV Lac ($L_X=10^{28.74}$~erg~s$^{-1}$) \cite{aud00}. 
Large flares releasing more than $5\times10^{32}$~erg in X-rays are observed in these 
objects at a rate of about one event per day \cite{aud00}. 

In impulsive solar flares, the ratio of the power contained in accelerated protons of 
kinetic energy above 10 MeV to the hard X-ray (1.6--12 keV) flare luminosity is estimated 
to be $L_p^\odot(\ge10~{\rm MeV})/L_X^{\rm hard}=0.09$ \cite{lee98}. With this value and 
the calculated Li production yield $Q({\rm Li}) = 2 \pm 1$~atoms per erg for $s=4$ (see 
Fig.~\ref{figvt1}), a flare that radiates more than $5\times10^{32}$~erg in X-rays can
produce more than $9\times10^{31}$ Li atoms in the star atmosphere. Detailed calculations 
for solar flares have shown that the bulk of the nuclear interactions of the flare-accelerated 
particles occur at atmospheric depths corresponding to column densities in the range 
$10^{-3}-10^{-1}$~g~cm$^{-2}$ \cite{hua89}. The models of Pavlenko et al. \cite{pav95} 
for the formation of Li lines in M-type dwarfs show that the Li I line at 6707.81~\AA~is 
also produced at depths $d \lsim 10^{-1}$~g~cm$^{-2}$. Thus, Li atoms can be 
synthesized in stellar flares right in the region where the strong Li I resonance 
line is formed. Taking the typical radius of mid-M type main-sequence dwarfs to be
$R=0.4~R_\odot$, the total number of H atoms in this region is 
$N_H \approx 4 \pi R^2 d/ (1.4 m_p) \lsim 4\times10^{44}$. The nucleosynthesis of 
more than $9\times10^{31}$ Li atoms in a large flare would thus give 
${\rm Li}/{\rm H}>2.2\times 10^{-13}$. From the models of Ref.~\cite{pav95} for
a star with $T_{\rm eff} = 3000$~K, surface gravity $\log g=5$ and a solar-like 
quiescient chromosphere, the corresponding equivalent width of the Li I $\lambda$6708 
line would be $EW({\rm Li})>80$~m\AA. But the Li production region can be significantly 
heated by the flare energy deposition, which could decrease the line strength (see 
\cite{pav95}).

Although challenging, the observation of Li enhancement during a strong flare of an active 
M-type dwarf appears to be possible. We have shown that energetic events occuring at a rate 
of about one per day could temporarily produce a Li absorption line of equivalent width 
$EW({\rm Li})\gsim80$~m\AA. This could be measured in high signal-to-noise spectra taken from 
nearby and relatively luminous objects, like AD Leo ($d=4.9$~pc; $V=9.43$~mag). Convective 
motions in the star upper atmosphere could rapidly reduce the line intensity. The detection 
of a variable Li absorption line from a flaring star would obviously shed new light on 
these dynamical processes. 

\section{Conclusions}
\label{secvt6}

The Li isotopes are likely to be the best tracers of the nucleosynthesis probably occuring in 
flares of active stars, because (i) background Li of protostellar origin can be rapidly depleted 
in cool dwarfs and (ii) $^6$Li and $^7$Li can be efficiently produced in stellar atmospheres by 
He+He reactions. If, as in solar flares, energetic $^3$He nuclei are strongly enriched by 
the particle acceleration process, $^6$Li can be mostly synthesized by the reaction 
$^4$He($^3$He,$p$)$^6$Li. The relatively high $^6$Li abundances recently measured in several
metal-poor halo stars near turnoff have to be confirmed after the work of Ref.~\cite{cay07}. In 
any case, we have shown that stellar flares could account for significant $^6$Li production in 
these objects, thus avoiding the need for a new pre-galactic source of this isotope, such as 
non-standard Big Bang nucleosynthesis and cosmological cosmic rays. 

We have proposed a model of Li production in flares that can explain the spread in 
Li abundances and the Li-activity correlation observed in young open clusters. Understanding
the origin of the Li dispersion is critical, because the Li abundance is considered to
be one of the best age indicator of young cool stars. Measurements of the isotopic ratio 
$^6$Li/$^7$Li in active stars of young open clusters could allow to test the flare 
production scenario. 

But the most direct evidence for flare nucleosynthesis would be the detection of Li enhancement 
during a strong stellar flare. We suggest that the best target for such an observation would be 
an active M-type dwarf in the solar neighbourhood, because (i) main-sequence M-type stars are 
almost free of background Li of protostellar origin and (ii) models for the formation of Li 
lines in very cool dwarfs show that Li abundance as little as a few times 10$^{-13}$ can be 
detected in these objects. We have estimated that large flares that could produce such a Li 
abundance may occur at a rate of about one per day in the most active M-type dwarfs. If the 
proposed observation is successful, it will certainly provide unique information on the nuclear 
processes occuring in stellar flares.

\vspace{0.5cm}
V.T. acknowledges financial support from the AGAUR (grant 2006-PIV-10044).

\end{document}